\documentclass[prd,superscriptaddress,nofootinbib,showkeys, preprint]{revtex4}
\usepackage{graphicx}
\usepackage[usenames,dvipsnames]{color}
\usepackage{mathtools}
\usepackage[colorlinks]{hyperref}
\usepackage{float}

\newcommand{\be}{\begin{eqnarray}}                                             
\newcommand{\ee}{\end{eqnarray}}
\newcommand{\nn}{\nonumber}

\begin{document}
	\title{    BRST symmetry as a mechanism for confinement in the inter gauge  framework}
	
	\author{Haresh Raval}
	\email{haresh$_$ein@yahoo.com} 
		\author{  Bhabani Prasad Mandal}
	\email{bhabani.mandal@gmail.com} 
	\affiliation{Department of Physics,  Institute of Science, Banaras Hindu University,
		Varanasi-221005, India}
		\begin{abstract}
			The finite field dependent BRST (FFBRST)  technique generally interrelates  two different gauge fixed theories. Here we propose a new and unique FFBRST  transformation that transforms the effective theory in Lorenz gauge, which represents perturbative or deconfined phase of QCD to massive gauge theory. The exercise serves two purposes. First is of course that it is a distinct connection where one of the theories is not gauge fixed and BRST invariant therefore, it is expected to give newer insights into FFBRST technique itself.  Second is that a particular case of this transformation holds a novel physical significance of acting as general mechanism for confinement across two different gauge theories. Thus, the present task  identifies a profound  non-perturbative implication of the BRST symmetry in a general setting.
		
	\end{abstract}
	\maketitle
	\section{Introduction}
	  The role of BRST transformation  in  quantizing the gauge theories is crucial. 
	The infinitesimal  anti-commuting  parameter of a usual BRST transformation can be generalized to be finite and field dependent  leaving the form of a transformation unchanged as was first
	done   in ref.~\cite{a12b}.  The explicit dependence on space time coordinates is absent in  the generalized parameter.    The field dependence of such FFBRST transformations varies the path integral
	measure keeping
	other characteristics  of 
	infinitesimal BRST transformations intact. Therefore,  the generating
	functional of a BRST invariant theory loses invariance
	under FFBRST. The change in the measure introduces a Jacobian in the path integral which depends on generalized parameter. Thus, the action that represents the Jacobian acts as a factor which can convert one effective theory to other. Therefore, FFBRST  may 
	be useful to get an insight into  phenomena in one theory given the knowledge of the same in the other theory. Hence, this technique finds  numerous applications in interrelating   two BRST invariant gauge fixed theories~\cite{a13b,a14b,a15b,a16b,a17b,a18b,a181b,a182b,a183b,a184b,a19b}.
	 
	 The dual superconductor picture in principle explains quark confinement through a linearly rising potential between static quarks~\cite{a20b,a21b}. This model relies on dominance of Abelian degrees of freedom of QCD. The proposal of Abelian dominance suggests that  Abelian degrees of freedom are sufficient to credibly express QCD at infrared energy scale \cite{a22b}. Off-diagonal gluons i.e.,  gluons not
	 associated with the Cartan subalgebra of the gauge group play an important role in the discussion of Abelian dominance. The
	 $SU(N)$ gauge theory has $N(N-1)$ off-diagonal
	 gluons. These gluons obtaining  large dynamical mass is
	 seen to offer the required evidence of  
	 Abelian dominance because in the non-perturbative limit, they tend to decouple from dynamics and therefore   diagonal
	 gluons survive as the only dominant degrees of freedom as they remain massless. Studies of Abelian dominance have been done in mostly in the Maximal Abelian gauge  which is a particular case
	 of an Abelian projection. Recently an exceptional quadratic gauge  $A^a_{\mu} ( x) A^{\mu a} ( x) = f^a ( x) ; \  \text{  for each $a$ }$ has been introduced which does not fall in the class of Abelian dominance. The theory in this quadratic gauge is shown to exhibit evidence of confinement in terms of Abelian dominance under condensation of the ghost fields \cite{a23b,a24b}. The ghost condensed or confined phase in this gauge is given as follows \cite{a23b}
	 \be\label{gc}
	\mathcal{S}_{GC} &=	&\int d^4 x\Big(- \frac{1}{4} F^a_{\mu \nu} F^{\mu \nu a}+ \frac{\zeta}{2} F^{a 2} + ( F^a + M_a^2)  A^a_{\mu} A^{\mu a}  \Big) 
	\ee
with  imaginary $M^2_a$ being given as 
e.g, for $SU(3)$, $M^2_3=M^2_8=0$. While for the off-diagonal gluons, $M^2_1=+im^2_1, M^2_2=-im^2_1, \  
M^2_4=+im^2_2,  M^2_5=-im^2_2, \  M^2_6=+im^2_3, M^2_7=-im^2_3 \ (m_1^2, m_2^2, m_3^2 $ 
are positive real$)$.   This is not BRST or gauge invariant because of the mass term.
The perturbative results of the QCD are best visible in the field theory of Lorenz gauge hence, the effective action of this gauge signifies the deconfined phase. It is given by 
	\be\label{11}\mathcal{S}_{eff} =\int d^4x [- \frac{1}{4} F^a_{\mu \nu} F^{\mu \nu a}
	  +\frac{\zeta}{2 } F^{a 2} +F^a  \partial_{\mu} A^{\mu a}- \overline{c^a}
	\partial^{\mu } ( D_{\mu} c)^a].  \ee
	It is invariant under the following usual  BRST transformation 
	\begin{eqnarray}\label{tra1}
	\begin{split}
	\delta c^d =& \ \frac{ g\omega}{2}f^{dbc}c^bc^c\\
	\delta \overline{c^d } = & \omega F^d \\
	\delta A^d_\mu=&\ \omega(D_\mu c)^d \\
	\delta F^d =&\ 0
	\end{split}
	\end{eqnarray}

Our aim in this letter is to find a single FFBRST transformation which can transform Lorenz gauge action in Eq.~\eqref{11} to the confined phase action in the quadratic gauge in Eq. \eqref{gc}. Therefore, the present field redefinition holds a profound physical relevance in   providing a  general mechanism which executes deconfinement to confinement phase transition across two different gauge theories. We have thus identified the general non-perturbative implication of the BRST symmetry itself. This work  generalizes the mechanism in ref. \cite{a25b}, where a field redefinition implements the phase transition but within the quadratic gauge only  to the inter gauge framework.

\section{Review of the FFBRST}

The usual BRST transformation which is characterised by infinitesimal, global, anti-commuting parameter is generalised to have the transformation parameter finite and field dependent with out affecting the underlying symmetry of the effective action \cite{a12b}.  In this section, we briefly review the essential steps to construct this FFBRST transformation. At first 
 a numerical parameter $\kappa \ (0\leq \kappa\leq 1) $ is introduced and all the fields are then made  $\kappa$ dependent
so that  $\phi (x,\kappa=0)=\phi(x) $ and $\phi (x,\kappa=1)=\phi^\prime(x) $, the transformed field.
The symbol $\phi$ is generic notation for  all the fields present in the theory.  The BRST transformation in Eq.~\eqref{tra1} is then given by
\be\label{infb}
d\phi = \delta_b[\phi(x,\kappa)]\Theta^\prime(\phi(x,\kappa))\ d\kappa
\ee
where $\Theta'$ is a finite field dependent anti-commuting parameter. It can include new fields that  do not constitute the BRST invariant action. The 
$\delta_b[\phi(x,\kappa)]$ is the form of the BRST transformation for the corresponding field as in 
Eq.~\eqref{tra1}. The
FFBRST is then developed by integrating Eq.~\eqref{infb} from $\kappa=0$ to $\kappa=1$ as~\cite{a12b}
\be\label{ffbrst}
\phi^\prime\equiv \phi(x,\kappa=1)=\phi (x,\kappa=0)+\delta_b[\phi(0)]\Theta [\phi(x)]
\ee
where $\Theta[\phi(x)] =\int_0^1 d\kappa^\prime\Theta^\prime[\phi(x,\kappa)] $.   Like usual BRST transformation, FFBRST
transformation leaves the effective action in Eq.~\eqref{11} invariant  but it
does not leave the path integral measure, ${\cal D}\phi $ invariant    since the transformation parameter is finite and field dependent.  It produces a non-trivial Jacobian  $J$ i.e., $\mathcal{D}\phi(\kappa) \rightarrow J(\kappa)\mathcal{D}\phi(\kappa)$. 
This $J$  can further be cast as a local exponential  functional of fields, $ e^{iS_J}$ within the path integral (where the $S_J$ is the action arizing from  the Jacobian factor $J$) 
if the following condition is satisfied~\cite{12}
\be \label{con}
\int { \cal D }\phi (x,\kappa) \left [\frac{1}{J}\frac{dJ}{d\kappa}-i\frac{dS_J}{d\kappa}\right ]e^{i(S_J+\mathcal{S}_{eff})}=0.
\ee
Thus the procedure for FFBRST may be summarised essentially in three steps as (i)  calculate the infinitesimal change in 
Jacobian, $\frac{1}{J}
\frac{dJ}{d\kappa} d\kappa $ using 
\begin{equation}\label{j}
\frac{J(\kappa)}{J(\kappa+d\kappa) }= 1-\frac{1}{J(\kappa)}\frac{dJ(\kappa)}{d\kappa}d\kappa
= \sum_\phi \pm \frac{\delta\phi(x,\kappa+d\kappa)}{\delta\phi(x,\kappa)}
\end{equation}
for infinitesimal BRST transformation, $+$ or $-$ sign is for Bosonic or Fermion nature of the field  $\phi$ respectively 
(ii)  make a suitable ansatz for $S_J$ by considering all possible terms, (iii)  then  check  Eq.~\eqref{con}
for this ansatz and if that is consistent, finally  replace $J(\kappa)$ by $e^{iS_J}$ within the functional integral in the expression of  the generating functional
\begin{equation}
W=\int {\cal D}\phi (x) e^{iS_{eff}(\phi)} \ \stackrel{\longrightarrow}{\small FFBRST} \ \int {\cal D}\phi (x,\kappa) J(\kappa)e^{iS_{eff}(\phi (x,\kappa))} = \int {\cal D}\phi (x) e^{iS_{eff}^\prime(\phi)}
\label{ww}
\end{equation}
Where the new effective action $ S^\prime_{eff}=S_J+S_{eff}$.
In the last step we have set   $\kappa=1$. 

\section{BRST symmetry as a general mechanism for confinement}
While  perturbative consequence of BRST symmetry in physical states and renormalization  of a theory is well acknowledged, its non-perturbative implication has just  been recently observed that too in restricted setting \cite{a25b}. The  passage through the generalized BRST to be constructed in this section has further significance in that it uncovers the same in a general setting namely, suitably generalized BRST symmetry executes deconfinement to confinement phase transition  across two different gauge fixed theories.  
To develop the required FFBRST transformation for the purpose, we start with Lorenz gauge action in Eq.~\eqref{11}, which signifies perturbative phase as our initial action. Since it is BRST invariant, the technique can be applied on it.  
	 Next we introduce the two sets of entirely new auxiliary fields $W^a, Y^a$ whose BRST transformations are unknown and  to be determined later. Now let us choose the following ansatz for the parameter
\be\label{th}
\Theta'[\phi(\kappa)]=-i  \int d^4 x \left [ \gamma_1\overline{c^d} \partial_\mu A^{\mu d} + \gamma_2 Y^d A^{\mu d} A_\mu^{ d}+ \gamma_3  W^d  A^{\mu d} ( D_{\mu} c)^d \right],
\ee
here $ \gamma_1, \gamma_2, \gamma_3$ are numbers related to FFBRST.  The field $W$ is taken as commuting scalar and the field $Y$ is considered anti-commuting scalar. Therefore, square of the parameter, $\Theta'^2=0$. Group index $d$ is summed over. The parameter in Eq.~\eqref{th} is distinct from the usual ones which are used in connection of two BRST invariant theories. It contains two new sets of fields of different nature  which do not constitute the action before FFBRST operation. We shall later see that they do not appear in the final action after FFBRST application also. They only appear  inside the parameter but has a subtle role in governing the physical phenomenon of phase transition.  It moreover has covariant derivative.

Due to this parameter, the change in the Jacobian under transformation in Eq.~\eqref{tra1} as per Eq.~\eqref{j} is given as follows
\be\label{j1}
\frac{1}{J}\frac{dJ}{d\kappa} &=& -\Big(-\frac{\delta \Theta'}{\delta \overline{c^d}}  F^d + \frac{\delta \Theta'}{\delta \partial^\mu A_\mu^d} \partial^\mu(D_\mu c)^d- \frac{\delta \Theta'}{\delta Y^d} \delta_b Y^d +\frac{\delta \Theta'}{\delta W^d} \delta_b W^d 
\nn\\&+&\frac{\delta \Theta'}{\delta A_\mu^d} (D_\mu c)^d- \frac{\delta \Theta' }{\delta c^d}\frac{g}{2}f^{def}c^ec^f- \frac{\delta \Theta'}{\delta \partial_\mu  c^d}\frac{g}{2}\partial_\mu (f^{def}c^ec^f) 
 \Big),\ \ \text{$d$ is summed over}\nn\\
&=&i\int d^4x\Big(- \gamma_1 F^d\partial_\mu A^{\mu d}+ \gamma_1 \overline{c^d} \partial_\mu D^{\mu} c^d- \gamma_2 \delta_b Y^d A^d_{\mu} A^{\mu d}+ \gamma_3\    A^{\mu d} ( D_{\mu} c)^d \delta_b W^d\nn\\&+&
    \Big[2\gamma_2 Y^d A^{\mu d}+\gamma_3\big(W^d    ( D^{\mu} c)^d -gf^{abc} W^a  A^{\mu a}\delta^{bd}c^c \big) \Big] (D_\mu c)^d \nn\\&+& \gamma_3 g^2  f^{abc} W^a  A^{\mu a}A_\mu^b \delta^{dc}  \frac{f^{def}c^ec^f}{2} \ -
\gamma_3  W^a  A^{\mu a} \delta^{ad}   \frac{ g}{2}\partial_\mu (f^{def}c^ec^f) \nn\Big)\ee
which after simplification of various terms leads to the following expression
\be
&=& i \int d^4x\Big( -\gamma_1 F^d\partial_\mu A^{\mu d} + \gamma_1 \overline{c^d} \partial_\mu D^{\mu} c^d  - \gamma_2 \delta_b  Y^d A^d_{\mu} A^{\mu d} + \gamma_3\   \delta_b W^d  A^{\mu d} ( D_{\mu} c)^d\nn\\&+&2  \gamma_2 Y^d  A^{\mu d} ( D_{\mu} c)^d +\gamma_3  W^d  (D^\mu c)^d    ( D_{\mu} c)^d  -  \gamma_3   W^d  A^{\mu d} \delta_b ( D_{\mu} c)^d \Big),
\ee	 where  $\delta_b$ represents BRST variation and we used identity 
\be
&&\delta_b ( D_{\mu} c)^d=  \frac{ g}{2}\partial_\mu (f^{def}c^ec^f)-     gf^{dbc}  ( D_{\mu} c)^b c^c -\frac{ g^2}{2} f^{dbc} A_\mu^b f^{cef}c^e c^f
.\ee 
Due to nilpotency, $ \delta_b ( D_{\mu} c)^d=0  \ \text{and additionally the identity} \ \  ( D^{\mu} c)^d  (D_\mu c)^d =0$, hence the last two terms vanish  in Eq.~\eqref{j1}. 

 The terms that have  $\Theta'$
as multiplicative
factor are not present in the $\frac{1}{J}\frac{dJ}{d\kappa} $, which suggests that  fields in the ansatz for the $S_J$ are $\kappa$ independent.
Therefore,  the action representing Jacobian, $S_J$ must be of the form 
\be
S_J[\phi(\kappa), \kappa] &=&  \int d^4x\Big( \alpha_1(\kappa)\   \delta_b W^d  A^{\mu d} ( D_{\mu} c)^d+  \alpha_2 (\kappa) \delta_b  Y^d A^d_{\mu} A^{\mu d}+  \alpha_3(\kappa) Y^d  A^{\mu d} ( D_{\mu} c)^d \nn\\ &+& \alpha_4(\kappa) F^d\partial_\mu A^{\mu d}+ \alpha_5(\kappa) \overline{c^d} \partial_\mu D^{\mu} c^d\Big)
\ee
where $\alpha_j(\kappa), j=1,\cdots,5$ are  functions with initial condition $\alpha_i(\kappa=0)=0$ and the fields are considered to be $\kappa$ independent.
Therefore, 
\be
i\frac{dS_J}{d\kappa}&=&i \int d^4x\Big(\dot{ \alpha}_1(\kappa)\   \delta_b W^d  A^{\mu d} ( D_{\mu} c)^d+  \dot{\alpha_2 }(\kappa) \delta_b  Y^d A^d_{\mu} A^{\mu d}+  \dot{\alpha_3}(\kappa) Y^d  A^{\mu d} ( D_{\mu} c)^d \nn\\ &+& \dot{\alpha_4}(\kappa) F^d\partial_\mu A^{\mu d}+ \dot{\alpha_5}(\kappa) \overline{c^d} \partial_\mu D^{\mu} c^d\Big),
\ee
where overdot represents $\frac{d}{d\kappa}$.
Condition in Eq.~\eqref{con} must hold which gives us  the following expression
\be
&&\int {\cal D}\phi[x,\kappa]\int d^4x\Big( [\dot{\alpha_1}(\kappa)-\gamma_3]\   \delta_b W^d  A^{\mu d} ( D_{\mu} c)^d  + [\dot{\alpha_2}(\kappa)+\gamma_2]  \delta_b  Y^d A^d_{\mu} A^{\mu d}\nn\\ &+&   [\dot{\alpha_3}(\kappa)-2\gamma_2]  Y^d  A^{\mu d} ( D_{\mu} c)^d + [\dot{\alpha_4}(\kappa)+\gamma_1]F^d\partial_\mu A^{\mu d}+ [\dot{\alpha_5}(\kappa)-\gamma_1] \overline{c^d} \partial_\mu D^{\mu} c^d\Big) \times\nn\\
&&e^{i(S_{eff}+S_J)}=0,
\ee
which yields the following relation among parameters
\begin{eqnarray}\label{tr}
\begin{split}
{\alpha_1}&=  \gamma_3 .\kappa\\
-{\alpha_2}&= \frac{1}{2}{\alpha_3}=\gamma_2.\kappa\\
{\alpha_4}&=-	{\alpha_5}=-\gamma_1.\kappa	\ .
\end{split}
\end{eqnarray}
We choose arbitrary parameters   $\gamma_1 =1, \gamma_2 =1$, 
$\gamma_3=2$. 
Thus, the additional Jacobian contribution at $\kappa=1$ is
\be
S_J &=& \int d^4x\Big( 2\   \delta_b W^d  A^{\mu d} ( D_{\mu} c)^d  -  \delta_b  Y^d A^d_{\mu} A^{\mu d}+ 2  Y^d  A^{\mu d} ( D_{\mu} c)^d -F^d\partial_\mu A^{\mu d}\nn\\ &+&    \overline{c^d} \partial_\mu D^{\mu} c^d\Big) 
\ee
Therefore, the resulting action at $\kappa=1$ can be given by
\be\label{im}
\mathcal{S}_{eff}+S_J 
&=	&\int d^4 x\Big[- \frac{1}{4} F^d_{\mu \nu} F^{\mu \nu d}+  2\   (\delta_b W^d+ Y^d)  A^{\mu d} ( D_{\mu} c)^d  -  \delta_b  Y^d A^d_{\mu} A^{\mu d} +\frac{\zeta}{2} F^{d 2}  \Big].	\ee	
Please note that $W^d$ does not appear in this final action, hence we are free to choose its transformation. 
We take 
\be \label{1star}
\delta_b W^d+ Y^d=0. \ee
With this transformation property, we get rid of undesired second term and also make sure that $Y^d$ disappear from the final action. Hence now we are at our will to choose its transformation as well. Thus we arrive at the action having the form of massive gauge theory upto a background auxiliary  term. A particularly interesting case arises when we take
\be\label{2star}
\delta_b Y^d = -(F^d + M_d^2),\ee 
where $M_d^2$ is set as elaborated below Eq.~\eqref{gc}. The transformations obey the following algebra
\be
\delta_b^3 W^d=0 \ \ \text{and}\ \  \delta_b^2 Y^d =0. 
\ee
With these transformations we have the following resulting action
\be
\mathcal{S}_{eff}+S_J 
&=	&\int d^4 x\Big[- \frac{1}{4} F^d_{\mu \nu} F^{\mu \nu d}+ \frac{\zeta}{2} F^{d 2} + ( F^d + M_d^2)  A^d_{\mu} A^{\mu d}  \Big]
.	\ee	
This is nothing but the action in the confined phase of the quadratic gauge. Thus we find a new general process of field redefinition (with parameter as in Eq.~\eqref{th} and transformations of intermediate auxiliary fields as in Eqs.~\eqref{1star}, \eqref{2star})  which can carry out deconfinement to confinement phase transition across two different gauge fixed theories. 
It is obvious but worth mentioning that the present   transition from deconfined phase in the Lorenz gauge to confined phase in the quadratic gauge can not be executed by the mechanism of ghost condensation.
\section{conclusion}
We proposed a unique  mechanism in redefinition of existing fields which executes deconfinement to confinement phase transition in the inter gauge framework namely, deconfined phase belongs to Lorenz gauge and confined phase to the quadratic gauge. Thus, the task classifies the non-perturbative implication of BRST symmetry in broader framework which  is generally  considered to be useful in just perturbative region.
 The order parameter of transition is given by  the FFBRST parameter  in this process.
 Both phases belong to two different gauge theories which compelled us to extend the FFBRST technique  to include covariant derivative  and two new sets of auxiliary fields, one of which is classical and the other is Grassmann field  in the field dependent parameter. These new auxiliary fields  disappear
 in the phase after FFBRST application but regulate the phase transition between two QCD phases through their transformation properties which  were unrestrained. Thus, this field redefinition turns out to be distinct from the usual ones.  The reverse of this phase transition i.e., from confined phase in quadratic gauge to deconfined phase in the Lorenz gauge can not be performed within the FFBRST technique as it works consistently only on BRST invariant actions.
\subsection*{{Appendix}}
\noindent Additional merit of  the given FFBRST structure lies in its flexibility. Let's select
\be \label{s1}
\delta_b W^d+ Y^d=-\overline{c^a} \ \  \text{and}\ \   \delta_b Y^d = -F^d \ee
keeping rest of the procedure same. It is interesting to observe that both of these transformations are nilpotent.
Then from Eq.~\eqref{im}, it can be deduced that
\be\label{a1}
\mathcal{S}_{eff}+S_J 
&=	&\int d^4 x\Big[- \frac{1}{4} F^d_{\mu \nu} F^{\mu \nu d}-  2\   \overline{c^a}  A^{\mu d} ( D_{\mu} c)^d  + F^d A^d_{\mu} A^{\mu d} +\frac{\zeta}{2} F^{d 2}  \Big].
\ee
Now, if \be \label{s2}
\delta_b W^d+ Y^d=-\overline{c^a} \ \  \text{and}\ \   \delta_b Y^d = \frac{1}{2\xi} A^d_{\nu} A^{\nu d} \ee
 then from Eq.~\eqref{im}, we have
 \be \label{a2}
 \mathcal{S}_{eff}+S_J 
 &=	&\int d^4 x\Big[- \frac{1}{4} F^d_{\mu \nu} F^{\mu \nu d}-  2\   \overline{c^a}  A^{\mu d} ( D_{\mu} c)^d - \frac{1}{2\xi} (A^d_{\mu} A^{\mu d})^2  \Big].
 \ee
 The term $F^{d 2}$ is dropped using its eq. of motion $F^d=0$ in this case.
 Both the actions in Eqs. \eqref{a1},\eqref{a2} are the effective actions in the quadratic gauge before ghost condensation. Therefore, the present FFBRST structure with transformations in Eq. \eqref{s1} or \eqref{s2}  additionally gives an alternate approach to the passage between the effective actions of Lorenz  and the quadratic gauges  discussed in spirit in ref.~\cite{a19b}. This example  proves for the first time that for a given passage between two theories, FFBRST transformation need not be lone. Hence, it is possible that some other FFBRST can also execute the phase transition elaborated in this paper which however is a matter of future investigation. 
 
{\bf Acknowledgements:} One of us (BPM) acknowledges the Research Grant for Faculty under IoE Scheme (number 6031).

 \end{document}